\documentclass[submit]{bioRxiv}

\usepackage{multirow}
\usepackage{rotating}
\usepackage{longtable}
\usepackage[switch]{lineno} 
\usepackage{setspace}
\linenumbers
\usepackage{tabularx}
\usepackage{float}
\makeatletter
\def\hlinewd#1{
\noalign{\ifnum0=`}\fi\hrule \@height #1 
\futurelet\reserved@a\@xhline}
\makeatother
\begin{document}
\nolinenumbers
\leadauthor{Vo}

\title{Two types of pyramidal cells and their role in temporal processing}
\shorttitle{Two types of pyramidal cells and their role in temporal processing}

\author[1, 2]{Anh Duong Vo}
\author[2]{Elisabeth Abs}
\author[1,2]{Pau Vilimelis Aceituno}
\author[$^*$,1,2]{Benjamin Friedrich Grewe}
\author[$^*$,3]{Katharina Anna Wilmes}
\affil[1]{ETH AI Center, ETH Zurich, OAT X11, Andreasstrasse 5, 8092 Zürich, Switzerland}
\affil[2]{Institute of Neuroinformatics, University of Zurich and ETH Zurich, Winterthurerstrasse 190, 8057 Zürich, Switzerland}
\affil[3]{University of Bern, Institute of Physiology, Bühlplatz 5, 3012 Bern, Switzerland}
\affil[$^*$]{jointly directed}
\date{}

\maketitle

\noindent Corresponding author: Katharina Anna Wilmes

\noindent Email address of corresponding authors: \href{mailto:katharina.wilmes@unibe.ch}{katharina.wilmes@unibe.ch} \\

\noindent Conflict of interest: The authors declare no competing financial interests.

\noindent Classification: Perspective


\newpage

\begin{abstract}
Recent work has provided new insights into the temporal specialization of Intratelencephalic (IT) and Pyramidal tract neurons (PT). However, functional and anatomical differences of IT and PT have not been connected yet. This perspective article contributes by highlighting empirical studies about the connectivity of IT and PT as well as their specialization in sensory and motor processing in diverse brain regions. We further review conceptual models that would align with the connectivity motif. We conclude that further research is needed to understand the role of the unidirectional connectivity from IT to PT in temporal processing and suggest concrete experiments.
\end{abstract}

\section{Introduction}
\begin{figure*}
    \centering
    \includegraphics[width=1\textwidth]{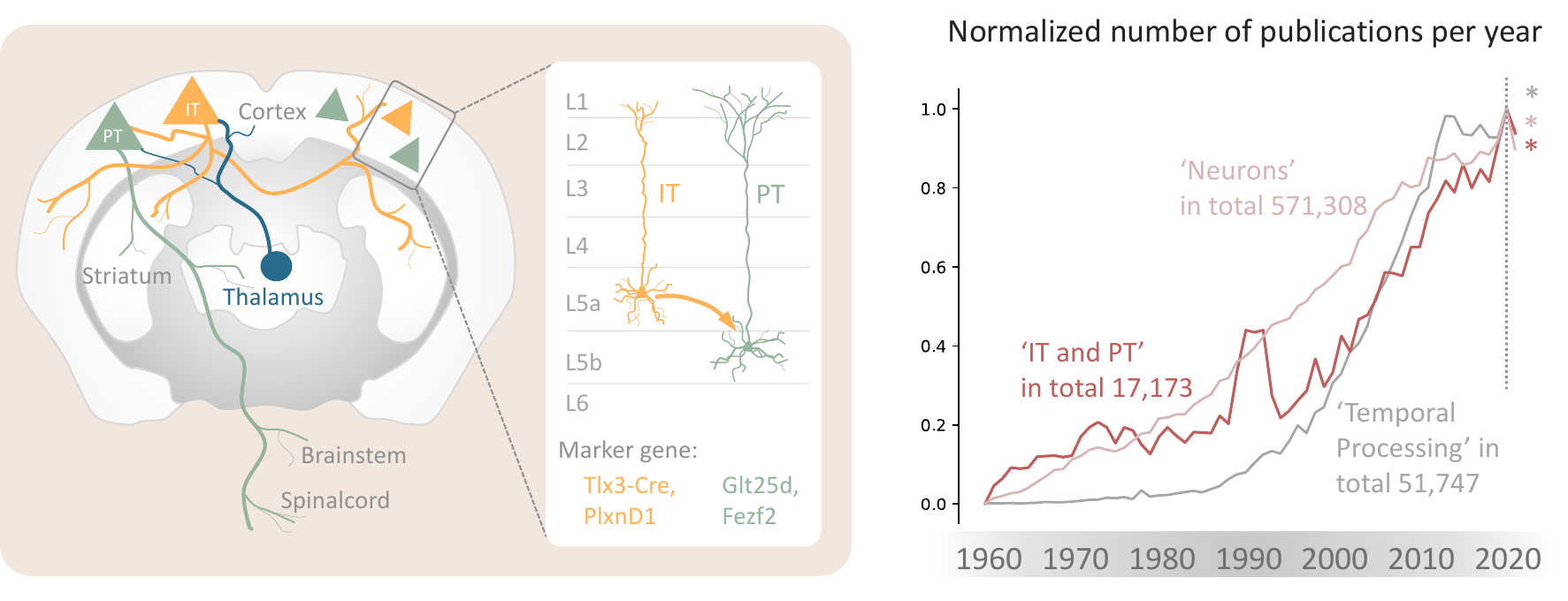}
    \caption{\textbf{Introduction to IT and PT.} Left: Illustration of projection areas of IT and PT and feedback loop from thalamus to cortical areas given a coronal section of the brain. Middle: IT and PT in cortical layers and morphological features. At the bottom, we list genes expressed by IT and PT used by \cite{Kim2015ThreeFunction.,Musall2023PyramidalDecision-making,Mohan2023CorticalSubnetworks}. Right: Based on the meta-analysis method introduced by \cite{Gurevitch2018Meta-analysisSynthesis}, we show the growing relevance of this research. A normalized number of publications per year and a total number of publications about the topics 'IT and PT' and 'Temporal Processing'. We added the statistics of publications about the topic 'Neurons' in general as a reference. The number of publications includes publications from 1961 till 2022, extracted from a Scopus search  (Status: July 2024). We normalized the data for each topic by its maximum value marked by the star (*). For the search, subject areas have been limited to Neuroscience, Medicine, Biochemistry, Genetics and Molecular Biology, and Psychology. We included different terms for IT and PT such as 'intratelencephalic', 'pyramidal tract', 'extratelencephalic', 'cortico-cortical', 'cortico-subcortical', 'L5a' and 'L5b' neurons. }
    \label{fig:introduction}
\end{figure*}

Recent advancements in cell classification techniques enabled a better understanding of brain circuits in terms of their connectivity, function, and development \citep{Zeng2017NeuronalForward}. This led to an increased understanding of the diverse functional roles of inhibitory cell types \citep{Fishell2020InterneuronControllers}, which was recognized in computational models of cortical circuits \citep{miller2020generalized,bos2020untangling,myers2022attentional,Kuchibhotla2017ParallelBehavior,Dipoppa2018VisionCortex,Wilmes2019InhibitoryRepresentations,Hertag2019AmplifyingTypes}. It has been known for a while that excitatory cells can be also distinguished into subtypes. In particular, layer 5 (L5) pyramidal cells, which can be distinguished into Intratelencephalic neurons (IT) and Pyramidal tract neurons (PT) based on their projection areas, have recently gained increasing attention (Fig.\ \ref{fig:introduction}) \citep{Harris2015TheVariations,Kawaguchi2017PyramidalCortex,Tasic2018SharedAreas,Peng2021MorphologicalTypes, Whitesell2021RegionalNetwork,Yao2023ABrain}. However, there are still many open questions about the different functionalities of excitatory cell types. Similar to inhibitory cell types, a deeper understanding of the roles of excitatory cell types could reveal novel insights behind the mechanisms in neural information processing. 

IT neurons play a key role in communication between cortical areas and the striatum, while PT neurons are responsible for connecting the neocortex to subcortical areas such as the brainstem, spinal cord, ipsilateral striatum, and thalamus \citep{Hirai2012SpecializedAreas,Kita2012TheRat,Ueta2014MultipleRats,Rojas-Piloni2017RelationshipsNeurons,Hattox2007LayerProperties,Groh2010Cell-TypeArea,Kim2015ThreeFunction.}. Studies have found differences in morphological and electrophysiological properties between IT and PT \citep{Baker2018SpecializedConsequences,Moberg2022NeocorticalBehavior,Shepherd2013CorticostriatalDisease}. PT have thick apical dendrites with distal branches that arborize in layer 1, whereas IT have thin apical dendrites with poorly branched distal branches (Fig.\ \ref{fig:introduction} (right)) \citep{Hattox2007LayerProperties,Groh2010Cell-TypeArea}. Additionally, PT show high-frequency bursts, while IT spike more regularly \citep{Hattox2007LayerProperties}. In experiments, PT and IT can be distinguished based on genetic markers and hence recorded, as well as, optogenetically stimulated and silenced in different mouse lines \citep{Baker2018SpecializedConsequences,Matho2021GeneticCortex}. These differences in connectivity, firing, morphology of the apical dendrites, and gene expression suggest task-related specializations between IT and PT. 

Interestingly, several experimental findings highlighted different specializations of IT and PT in temporal processing \citep{Lur2016Projection-SpecificSubnetworks,Kim2015ThreeFunction.,Bae2021ParallelNeurons,Dembrow2015TemporalInputs, Rindner2022Cell-type-specificCortex, Li2015AMovement}. \cite{Li2015AMovement}, \cite{Lur2016Projection-SpecificSubnetworks} and \cite{Moberg2022NeocorticalBehavior} even mentioned that there might be a link between the unidirectional connectivity of IT/PT and information processing. However, a thorough evaluation of the relationship between connectivity and the different roles of IT and PT is still missing. Here, we review the different roles of IT and PT in L5 in temporal processing and explore how the IT-PT circuit motif might enable the processing of information. 

We begin by summarizing the connectivity motifs of IT and PT in L5, where we highlight the unidirectional connection from IT to PT. We review recent studies of IT and PT related to their distinct roles in temporal processing in sensory discrimination and motor tasks. We continue by proposing models of temporal processing that would align with the unidirectional connectivity from IT to PT and their distinct roles. To offer a more complete perspective, we discuss studies that challenge the proposed conceptual models. Finally, we suggest potential future experiments to test the functional relevance of the canonical IT-PT circuit motif for temporal processing.

\section{Connectivity motifs of IT and PT} \label{sec:connectivity}

\begin{figure*}
    \centering
    \includegraphics[width=0.7\textwidth]{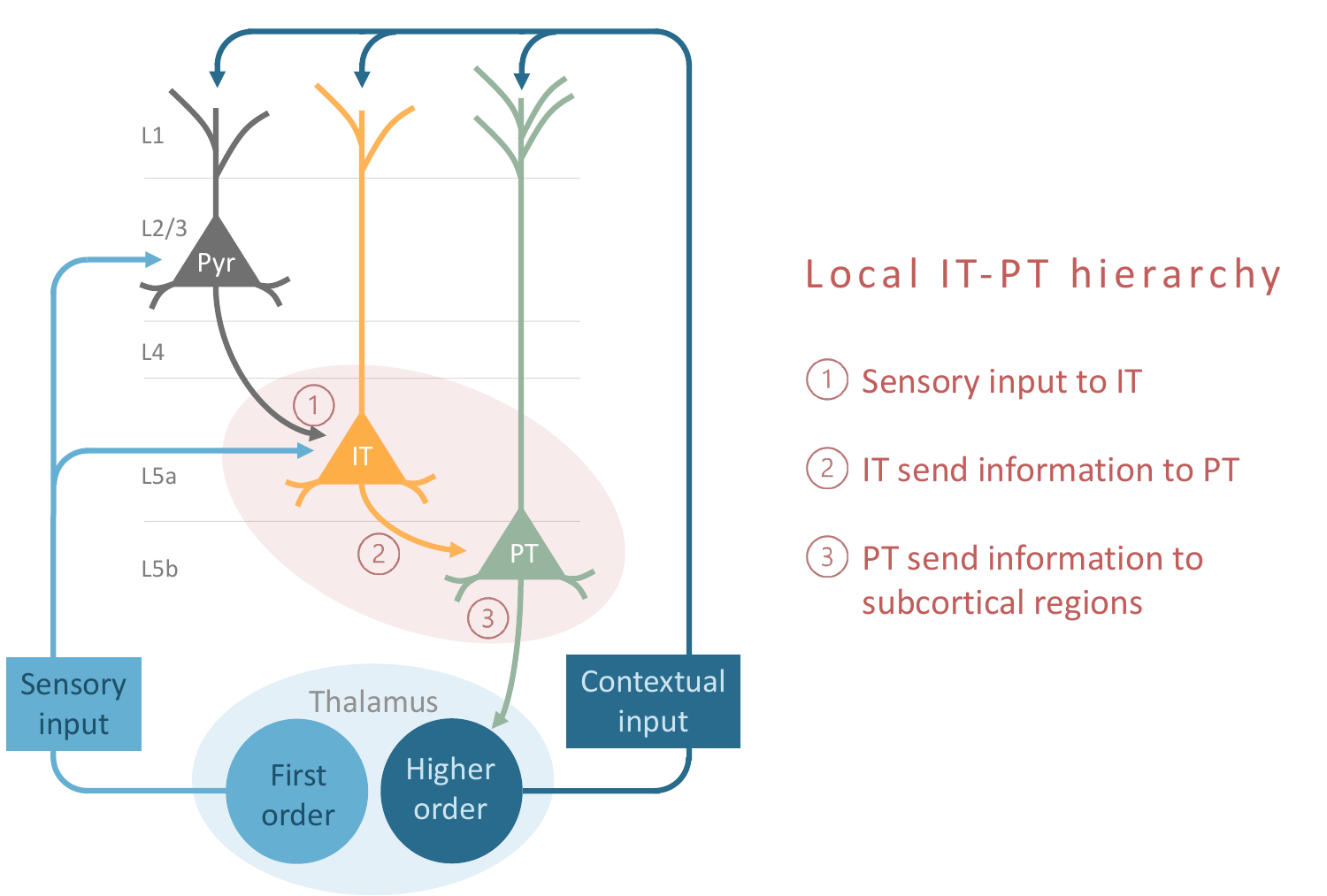}
    \caption{\textbf{Connectivity of IT and PT.} Connections between pyramidal cells (Pyr) in L2/3 and 5 and thalamus. For simplicity, we only included the connections of interest.}
    \label{fig:functional_connectivity}
\end{figure*}

Investigating the connectivity patterns of different neuronal subtypes is crucial to understanding their roles in complex brain functions. IT and PT are distinguished not only by their projection targets but also by their connection to each other.

Studies have revealed that IT excite PT but not vice versa. This unidirectional connection has been experimentally demonstrated in several cortical areas, including visual cortex \citep{Campagnola2022LocalNeocortex,Brown2009IntracorticalTargets}, frontal cortex \citep{Morishima2006RecurrentCortex}, and motor cortex \citep{Kiritani2012HierarchicalCortex}. The experimentally measured connection probabilities within IT, PT, and between them can be found in Table \ref{tab:connectivity}. Various sources have stated that this connectivity motif might support a local hierarchy between IT and PT \citep{Harris2015TheVariations,Moberg2022NeocorticalBehavior,Hooks2013OrganizationCortex,Harris2019HierarchicalConnectivity}. \cite{Moberg2022NeocorticalBehavior} even described the concept that IT may maintain and update cortical neurons for more efficient computations and PT broadcast this updated information to subcortical structures. 

This hierarchical structure is further supported by the input to IT and PT. Both neural populations receive input from the thalamus \citep{Kim2015ThreeFunction.,Yao2023Long-rangeCortex}. However, cortical IT receive stronger input from the thalamus and cortical areas compared to PT. This can be interpreted as a loop between the thalamus, IT and PT which was mentioned by \cite{Shepherd2021UntanglingPuzzle} (Fig.\ \ref{fig:functional_connectivity}).

To conclude, the unidirectional connection between IT and PT might support a local hierarchy between the two distinct neural populations. 

\section{Roles of IT and PT Neurons in Temporal Processing} \label{sec:temp}

\begin{figure*}
    \centering
    \includegraphics[width=0.9\textwidth]{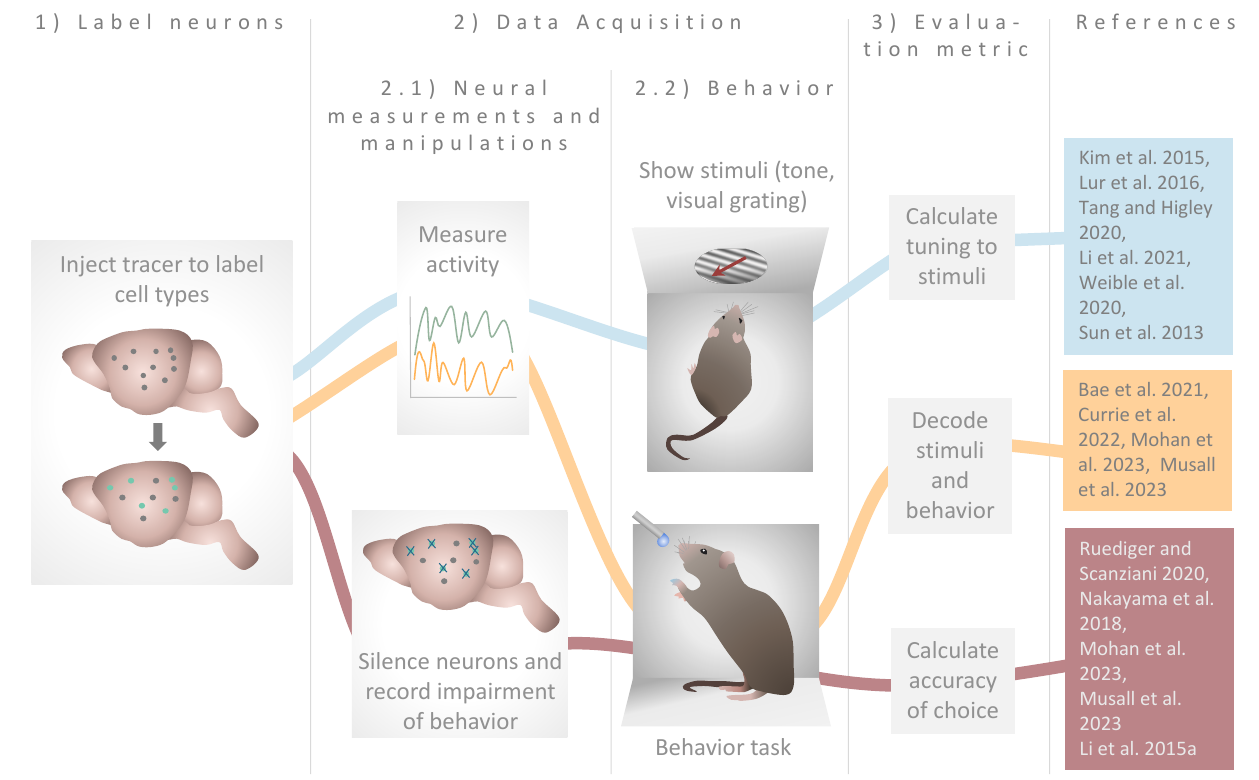}
    \caption{\textbf{Experiments related to information processing by IT and PT.} Here we describe different types of data acquisition and evaluation metrics to assess the features of different cell types.}
    \label{fig:experiments}
\end{figure*}

Given the distinct connectivity of IT and PT, it is not surprising that they specialize differently \citep{Baker2018SpecializedConsequences,Lesicko2022DiversePathway}, especially in disease \citep{Shepherd2013CorticostriatalDisease} and behavior \citep{Moberg2022NeocorticalBehavior}. Experimental findings further suggest that these cell types contribute to the ability to process stimuli in time --- from the perception of the duration of events to the processing of timing and direction of stimuli. This fundamental ability of animals, also referred to as temporal processing, has been extensively studied and shown to be a critical component of sensory and motor experiences \citep{Paton2018TheFunctions,Tsao2022TheDurations, Eichenbaum2017OnMemory,Teki2017TheMemory,Tallot2020NeuralBrain,Issa2020NavigatingTime,Petter2018IntegratingLearning,Mioni2020UnderstandingStudies,Golesorkhi2021TheProcessing,Wolff2022IntrinsicSegregation}. 

We explore the roles of IT and PT in processing information, specifically focusing on temporal aspects. We discuss their involvement in two main types of tasks: \textit{sensory discrimination tasks} and \textit{motor tasks}. We define sensory discrimination tasks as tasks that require animals to distinguish between sensory inputs, such as identifying visual patterns and the direction of moving patterns or differentiating between sounds. We further define motor tasks as tasks that involve motor actions by the animal directed towards a target. Some tasks might involve recognizing appropriate sensory cues before a motor action is taken which could be considered as both a sensory discrimination task and a motor task. However, we classify tasks as motor tasks if the primary requirement is motor activity, even though sensory decision-making is involved. If an experiment primarily assesses the differentiation between sensory patterns, we categorize it as a sensory discrimination task.

In this review, the terms \textit{first-order processing} and \textit{second-order processing} are used to describe the specializations of IT and PT. We define first-order processing as an essential step that lays the groundwork for subsequent neural processing stages, referred to as second-order processing. These terms are inherently relative, as any processing stage may act as either a first or second-order process depending on its contextual role in the sequence. For this review, we focus specifically on the interrelations between the specializations within IT and PT, analyzing the sequential order of these processing stages if they are interconnected.

\subsection{Sensory discrimination tasks}
\begin{figure*}
    \centering
    \includegraphics[width=1\textwidth]{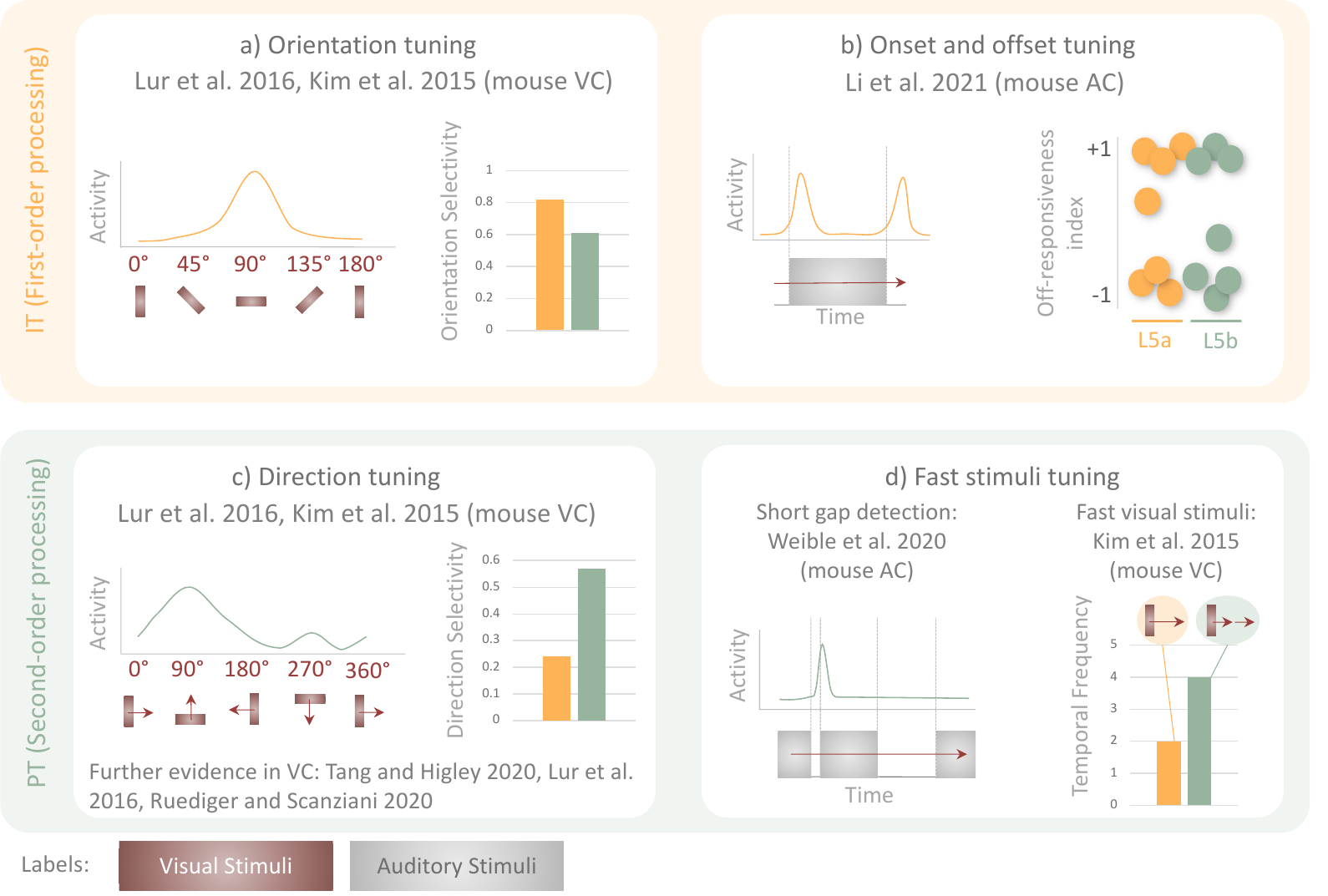}
    \caption{\textbf{Specialization of IT and PT in sensory discrimination tasks.} Summary of literature about the roles of IT and PT in sensory discrimination tasks with visual (dark red) and auditory stimuli (olive). Yellow arrows mark the connection from IT to PT. \textbf{Functionalities of IT are a) Orientation tuning:} (Left) Illustrative abstract based on \cite{Lur2016Projection-SpecificSubnetworks, Kim2015ThreeFunction.} shows that IT are tuned to orientation. (Right) Replotted data from \cite{Kim2015ThreeFunction.} describes the median orientation selectivity of IT and PT. IT are more orientation selective than PT; \textbf{b)  Onset and offset tuning:} (Left) Illustrative abstract based on \cite{Li2021PhasicDuration} shows that IT are tuned to onset and offset of tones. (Right) Illustration of experimental results of \cite{Li2021PhasicDuration}. The figure shows the off-responsiveness index (ORI) of L5a and L5b pyramidal neurons. ORI describes the relative strength of onset and offset responses in an individual cell, with ORI = 1 indicating Off-responsiveness only and ORI = 1 indicating On-responsiveness only. IT and PT are tuned to tone onset and offset. \textbf{Functionalities of PT are c) Direction tuning:} (Left) Illustrative abstract based on \cite{Lur2016Projection-SpecificSubnetworks,Kim2015ThreeFunction.} shows that PT are tuned to direction. (Right) Replotted data from \cite{Kim2015ThreeFunction.} describes the median direction selectivity of IT and PT. PT are more direction selective than IT; \textbf{d) Fast stimuli tuning:} (Left) Short gap detection: Illustrative abstract based on \cite{Weible2020ADetection} shows that PT is tuned to short gaps in tone. (Right) Fast visual stimuli: Replotted data from \cite{Kim2015ThreeFunction.} describes the median temporal frequency of IT and PT. PT prefer faster stimuli than IT. More information about the figures can be found in Table \ref{tab:figure_data}.}
    \label{fig:sensory_discrimination_abstract}
\end{figure*}

Sensory discrimination tasks refer to tasks in which an animal has to distinguish different directions of moving stimuli or the durations of stimuli. In the following, we summarize the literature on the specialization of IT and PT in sensory discrimination tasks.

\subsubsection{Direction discrimination}\label{sec:direction}  

Experiments involving animals' discrimination of different angles of moving gratings have shed light on visual processing. To determine the direction of movement, the brain must process a continuous stream of information. In the following, we describe how IT and PT contribute to direction discrimination. 

By comparing visual feature extraction across L5 populations, \cite{Lur2016Projection-SpecificSubnetworks} observed that PT exhibited broader orientation and spatial frequency tuning than IT in the visual cortex. In a more recent study by the same group, they showed that the visual cortex encodes sensory and motor information during a visually cued eyeblink conditioning task \citep{Tang2020LayerBehavior}. They confirmed again that PT are more broadly tuned towards visual stimuli. Based on this feature, \cite{Lur2016Projection-SpecificSubnetworks} referred to PT as complex cells which are characterized by being less sensitive to the exact location of the stimulus within their receptive field and respond best to moving stimuli. They further referred to IT as simple cells that are characterized by being more narrowly tuned and more sensitive to the orientation of lines or edges within their receptive field \cite{Hubel1962ReceptiveCortex}. \cite{Kim2015ThreeFunction.} confirmed the complex-cell and simple-cell-like features of PT and IT. They classified the cell types IT and PT according to their projection target. By showing moving gratings to the animals, they measured the orientation and direction selectivity of those cell types. While IT were highly selective to orientation, PT were highly selective to direction. Further, PT's high direction selectivity extends to low contrast settings \citep{Ruediger2020LearningCortex, Lur2016Projection-SpecificSubnetworks}.

Thus, IT encode the orientation of visual stimuli, which can be considered as first-order processing. PT encode the direction of visual stimuli which can be considered as second-order processing since it requires the integration of the orientation of visual stimuli over time.

\subsubsection{Detection of fast stimuli} 
The ability to detect and respond quickly to changes in sensory input is essential for survival. This process relies on sensory neurons that can rapidly process and transmit information. In the following, we explain the specialization of IT and PT in the detection of fast stimuli.

The work mentioned in Section \ref{sec:direction} by \cite{Kim2015ThreeFunction.} came to an important conclusion in terms of fast visual input. They varied the temporal frequency of the visual stimulus and observed that PT prefer faster stimuli than IT in the visual cortex. This is consistent with the literature on auditory gap detection. Gap detection describes the ability to detect short gaps in ongoing tones \citep{Rajasingam2021TheModulation}. \cite{Weible2020ADetection} aimed to understand why the auditory cortex is necessary to detect short gaps, but not long gaps. They identified a cortico-collicular connection, a connection by a subtype of PT, in the auditory cortex that amplifies cortical gap termination responses. They found that if onset and offset responses are temporally close to each other (short gap), the responses amplified each other in the auditory cortex and were detected as one strong event. If onset and offset are temporally far away from each other (long gap), the responses were detected as two separate and less strong responses. Interestingly, \cite{Li2021PhasicDuration} found that L5a neurons in the auditory cortex, which are most likely IT neurons, are selective for the onset and offset of tones. Further, IT demonstrate sharper tuning in tone frequency compared to PT in the auditory cortex \citep{Sun2013Synaptic5}.

To sum up, PT become more selective to fast stimuli. Interestingly, IT encode orientation, tone frequency, and stimulus onset and offset. Considering both specializations, the role of PT can be interpreted as second-order processing since selectivity to fast stimuli builds on the information of orientation, tone frequency, and stimulus onset and offset which would be first-order processing by IT.

\subsection{Motor tasks}
\begin{figure*}
    \centering
    \includegraphics[width=1\textwidth]{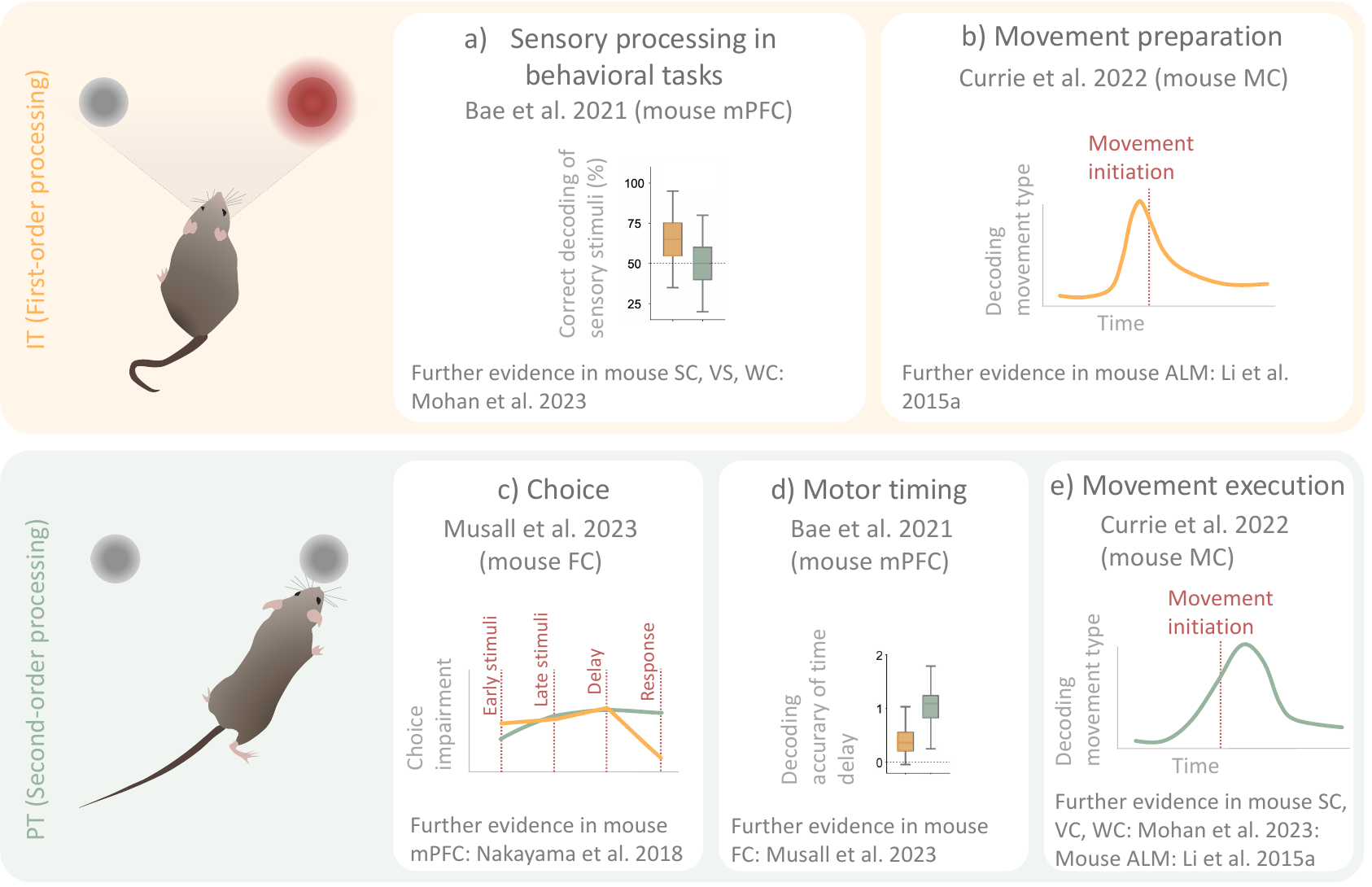}
    \caption{\textbf{Specialization of IT and PT in motor tasks.} Summary of literature about how IT and PT contribute to motor tasks. The yellow arrow marks the connection from IT to PT. \textbf{Specialization of IT are a) Sensory processing:} Adapted with permission from \cite{Bae2021ParallelNeurons}. The figure shows the correct decoding accuracy of sensory stimuli by IT and PT. IT encode visual stimuli better compared to PT; \textbf{b) Movement preparation:} Adapted visual abstract with permission from \cite{Currie2022Movement-specificClasses}. The figure shows the ability to decode movement type by IT through time. IT encode movement type better before movement initiation than after. \textbf{Specialization of PT are c) Choice:} Illustration of experimental results of \cite{Musall2023PyramidalDecision-making} describes impairment of choice by silencing IT and PT during the experiment. Silencing PT impaired choice stronger during choice compared to silencing IT; \textbf{d) Motor timing:} Adapted with permission from \cite{Bae2021ParallelNeurons}. The figure shows the decoding accuracy of delay time by IT and PT. PT encode delay time better compared to IT; \textbf{e) Movement execution:} Adapted visual abstract with permission from \cite{Currie2022Movement-specificClasses}. The figure shows the ability to decode movement type by PT through time. PT decode movement type better after movement initiation than before. More information about figures can be found in Table \ref{tab:figure_data}.}
    \label{fig:motor_tasks_abstract}
\end{figure*}

Motor skills are crucial for controlling and coordinating movement in the nervous system. They enable the brain to initiate and execute movements with temporal precision. Typically, a motor task involves presenting sensory cues to animals, to which they respond through specific motor actions. It can also involve anticipating an event. In the following section, we describe behavioral studies that analyzed specializations of IT and PT in motor tasks.

\subsubsection{Sensory processing in behavioral tasks}
To solve motor tasks, animals process sensory input which they use to collect information for their behavioral decision. In the following, we describe different behavioral studies that emphasize the different specializations of neurons in sensory processing before movement execution.

\cite{Musall2023PyramidalDecision-making} recorded the activity of different cell types in the auditory, parietal, and frontal cortices of mice during auditory decision-making. In this experiment, mice were presented with click sequences on both sides during a sampling phase. After a delay phase, if the mice chose the spout where more clicks were presented, they received a reward. A decoding model on the activity of all observed brain areas showed that IT encoded stimuli better than PT, while PT encoded choice better than IT. This was also supported by the variance of activity for different stimuli and choices. To causally test the functional role of IT and PT, they performed type-specific optogenetic inactivation during different phases: early audio, late audio, delay, and response. Silencing IT in frontal cortices impaired performance during early audio, late audio, and delay, while silencing PT in frontal cortices impaired performance largely during late audio and delay, but less during early audio. This suggests that IT contribute more to general stimuli encoding compared to PT in the frontal cortex.

Similar to auditory, parietal, and frontal cortices, IT respond to sensory stimulation in the somatosensory cortex \citep{Mohan2023CorticalSubnetworks}. They observed that stimulation of whiskers and the orofacial region strongly activates IT more than PT in the primary visual, whisker, and mouth–nose somatosensory cortex. Thus, it implies that IT is more involved in sensory processing than PT specifically in those brain areas.

The sensory processing specialization of IT is supported by \cite{Bae2021ParallelNeurons} who conducted a working memory experiment in the medial prefrontal cortex. During a match-to-sample task, mice were shown a binary sensory cue during the initial sampling phase, which indicated the solution to the task. A second sensory cue was presented after a random delay, followed by the mice choosing their movement and receiving a reward or not. The activity of IT during the delay phase encoded more information about the sensory cue in the sample phase than PT. Thus, IT contribute more to the processing of sensory stimuli than PT. It should be noted that in this task, next to sensory processing, also working memory is involved.

\subsubsection{Motor timing}
Animals can anticipate events and execute movements at the right moment, referred to as motor timing.

\cite{Bae2021ParallelNeurons} found that the activity of PT during the delay phase encoded more information about the temporal length of the delay than IT which was classified as ramping activity (Fig.\ \ref{fig:motor_tasks_abstract}). \cite{Musall2023PyramidalDecision-making} also observed increasing activity of PT before the movement execution. The observed ramping activity of PT is consistent with previous ramping models in motor timing \cite{Paton2018TheFunctions}, indicating that PT fulfill a timing role in this behavioral task. 
Thus, PT carry more temporal information compared to IT and are specialized in motor timing.

\subsubsection{Choice}
During a behavioral task, animals need to make choices to receive a reward or avoid punishment. Here we describe different behavioral studies that evaluated the specialization of IT and PT.

PT and IT contribute differently to decision-making \citep{Musall2023PyramidalDecision-making}. Silencing PT in the frontal cortex during the choice phase led to a significant impairment in performance while silencing IT during the same phase had no significant effect. This implies that PT contribute more to choice than IT.

\cite{Nakayama2018Cell-Type-SpecificBehaviors} came to similar conclusions. They studied behavioral flexibility and impulse control by training mice to perform a binary choice task that requires frequent adjustment of behavior. They found that inhibiting PT led to greater impairment in adjusting the behavior compared to inhibiting IT, supporting the idea that PT influence choice more than IT. In conclusion, PT contribute to choice in motor tasks.\\

We want to further evaluate the role of IT and PT in sensory processing in behavioral tasks, motor timing, and choice in the context of first and second-order processing. Motor timing and choice build on processing a sensory cue successfully to trigger an appropriate action. Therefore, sensory processing can be interpreted as first-order processing whereas motor timing and choice can be interpreted as second-order processing. Considering that IT specialize more in sensory processing in behavioral tasks and PT specialize more in motor timing and choice, IT might be more involved in first-order processing and PT might be more involved in second-order processing.

\subsubsection{Motor preparation and execution}
The preparation of the motor execution and the motor execution itself are further crucial aspects of motor tasks. In the following, we describe a behavioral study that evaluates the different roles of IT and PT in motor preparation and execution.

\cite{Li2015AMovement} recorded neuronal activity in the motor cortex of mice during a whisker-based object location discrimination task. They used optogenetics to selectively activate or inhibit IT and PT and observed the effects on motor planning and execution. They found that IT showed more distributed preparatory activity before upcoming movements rather than directly triggering specific movements. PT exhibited a strong bias during late movement planning, specifically when preparing for contralateral movements. This suggests that PT are more involved in triggering the movement execution. Work by \cite{Currie2022Movement-specificClasses} confirmed the results. They analyzed whether IT or PT in the motor cortex encode the type of movement during a pull and push task. The mice had to push or pull a horizontal lever during the presentation of an auditory cue to receive water. They found that IT encoded the movement type before initiation while PT encoded it during execution. \cite{Mohan2023CorticalSubnetworks} further showed that when PT were silenced during movement execution, the animals’ ability to lick and move their forelimbs decreased.

Thus, IT are more specialized in motor preparation which can interpreted as first-order processing. PT are more specialized in motor execution processing which can be interpreted as second-order processing as it follows the motor preparation.\\

In summary, IT and PT contribute in different ways to information processing in sensory discrimination and motor tasks. In sensory discrimination tasks, IT specialize in processing the orientation of sensory stimuli as well as onset and offset and frequency of auditory stimuli (first-order processing). PT specialize in processing temporal detection of direction and fast stimuli (second-order processing). In motor tasks, IT specialize in processing sensory input and motor preparation (first-order processing) while PT specialize in motor timing, choice, and motor execution (second-order processing).\\

\section{From connectivity to function}
\begin{figure*}
    \centering
    \includegraphics[width=1\textwidth]{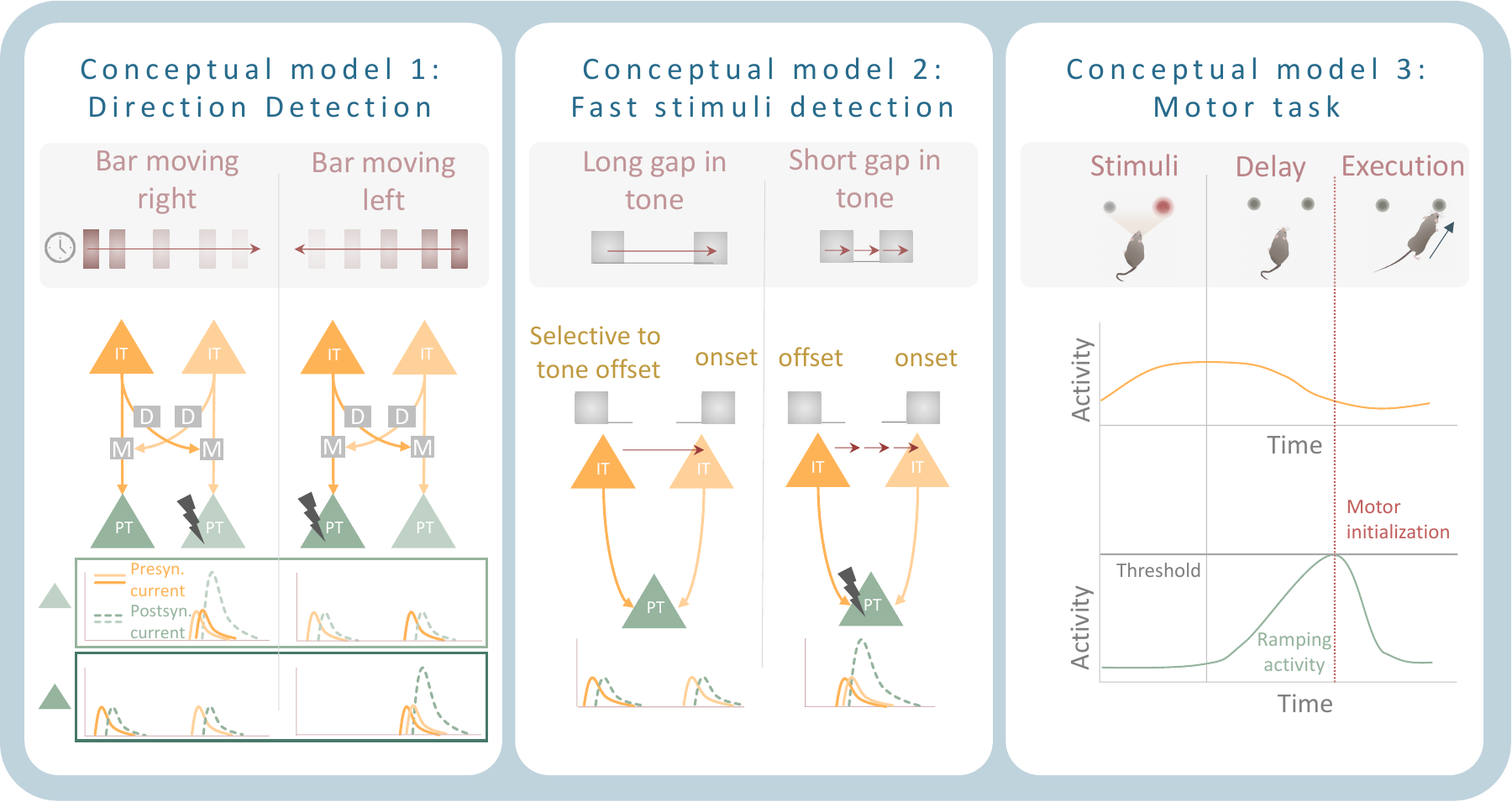}
    \caption{\textbf{Illustration of proposed conceptual models. Conceptual model 1)} Direction detection model including different cell types: the direction detection model is based on the Hassenstein-Reichardt model \citep{Hassenstein1956SystemtheoretischeChlorophanus,Yan2021TheModel,Mauss2017VisualSelectivity}. PT receive sensory information from orientation selective IT. (Left) For the bar moving to the right, temporal delay (D) and multiplication (M) between IT and PT enable amplification of activity in the right PT neuron. (Right) For the bar moving to the left, temporal delay and multiplication between IT and PT enable amplification of activity in the left PT neuron. \textbf{2)} Fast stimuli detection model including different cell types: A similar mechanism has been explored by \cite{Weible2020ADetection}. PT neuron receives information from onset and offset selective IT. (Left) For long gaps, the left and the right IT are activated with a longer time delay leading to no amplification of the activity of the PT neuron. (Right) For short gaps, left and right IT are activated with only a short time delay leading to an amplification of the activity of the PT neuron. \textbf{3)} Motor task model including different cell types: This is a ramping activity model reviewed by \cite{Paton2018TheFunctions}. IT which are responsible for sensory processing during the stimuli and delay phase send information to PT. PT exhibit ramping activity before movement initiation based on information from IT. Ramping activity of PT has been seen in \cite{Bae2021ParallelNeurons}.}
    \label{fig:concmodels}
\end{figure*}

The reviewed studies above have shown that IT encode first-order features, while PT encode second-order features. Interestingly, there exists a unidirectional connection from IT to PT (Section \ref{sec:connectivity}). The unidirectional connection may allow IT to send preprocessed first-order information to PT, thereby enabling PT to process second-order features. In the following, we describe conceptual models of direction discrimination, fast stimuli detection, and motor timing, each incorporating unidirectional connections (Table \ref{tab:conceptual_models}, Fig.\ \ref{fig:concmodels}). 

\subsection{Sensory discrimination task model}
In this section, we conceptualize models for sensory discrimination tasks including IT and PT. 

\subsubsection*{Conceptual Model 1: Direction detection}
We start by describing a model for direction discrimination in which the cells are selective to distinct directions of moving stimuli. 

Three main theories explain the processing of direction in the cortex according to \cite{Lien2018CorticalSynapses}. First, it is believed that cortical tuning arises from thalamic input \citep{Lien2018CorticalSynapses,Stanley2012VisualSynchrony,Li2013IntracorticalCortex,Lien2013TunedCircuits}. More specifically, cortical direction selectivity emerges at convergence of thalamic synapses. Second, other studies showed that retinal input to the thalamus is already direction-tuned \citep{Rasmussen2020ContributionsProcessing,Hillier2017CausalCortex,Kerschensteiner2022FeatureCells,Rasmussen2020ASelectivity,Bereshpolova2019ActivationNeuron}. Third, intracortical models propose that recurrent connections between neurons result in a spatial offset between excitation and inhibition or between excitatory inputs with distinct time courses \citep{Aponte2021RecurrentCortex,Li2015StrengtheningCortex,Freeman2021ACortex}. Although all three mechanisms could simultaneously contribute to selectivity, the IT-PT circuit motif supports the latter theory, as we will explain in the following. 

Literature has shown that IT are more tuned to the orientation while PT are more tuned to the direction of moving gratings. Based on the unidirectional connectivity from IT to PT, we hypothesize the following information flow: IT receive and process first-order features such as orientation and tone frequency, and then forward that information to PT. PT would use this information to detect second-order features such as the direction of the sensory input. To illustrate this, we mapped IT and PT to the well-known Hassenstein-Reichhardt direction selectivity model \citep{Hassenstein1956SystemtheoretischeChlorophanus,Yan2021TheModel,Mauss2017VisualSelectivity} (Fig.\ \ref{fig:concmodels}.1). The multiplicative operation in the model can be implemented by neuronal gain modulation \citep{Ferguson2020MechanismsCortex} or dendritic nonlinearities \citep{Poirazi2020IlluminatingModels}. 

\subsubsection*{Conceptual model 2: Fast stimuli detection}
In the following, we describe our model for fast stimuli detection. Fast stimuli detection is a subtask of duration detection in which the cells should be more responsive to fast-moving stimuli. We propose two different models for auditory and visual stimuli.

Previous research has also shown that the spatial structure of sensory inputs processed by orientation or frequency-tuned neurons plays a role in duration detection \citep{Froemke2013Long-termPerception,Gilbert2009PerceptualPlasticity,Karmarkar2007TimingStates}. Considering that IT is highly orientation and frequency selective, the connection between IT and PT might be relevant to help PT become more selective to fast stimuli.

For auditory stimuli, we propose conceptual model 2 for fast stimuli detection (Fig.\ \ref{fig:concmodels}.2). We focus on the task of short gap detection in tone which has been explored as a specialization of PT in \cite{Weible2020ADetection}. Here, PT are more responsive to short gaps in tone. The model that we propose is based on experimental findings of the same authors who showed that fast stimuli responses are amplified due to short delays between tone onset and offset responses. In this model, we want to leverage that IT are tuned to onset and offset of auditory stimuli \citep{Li2021PhasicDuration}. The unidirectional connectivity might allow IT to send information about the onset and offset of the stimuli to PT. Given a large gap in tone, PT would receive the onset and offset responses with a large delay which would not lead to any amplification of response. Given a short gap in tone, PT would receive the onset and offset responses with a short delay which would lead to any amplification of responses, since the onset and offset responses happen almost at the same time.

For visual stimuli, we propose a similar model. Detecting fast-moving visual stimuli has been studied as a specialization of PT by \cite{Kim2015ThreeFunction.}. For visual stimuli, we propose a similar mechanism as for conceptual model 2 in which IT send information about the onset and offset of visual bars to PT while PT experience either no amplification of response for slowly moving bars or an amplification of response for fast moving bars. There is no experimental evidence for visual onset and offset selectivity of IT. However, it has been shown that IT is highly orientation selective \citep{Kim2015ThreeFunction.} which is a base for onset and offset selectivity. \\

We proposed conceptual models for direction detection and fast stimuli detection. The models exploit the unidirectional IT-PT connection that enables IT to send the processed first-order features such as orientation or onset and offset information to PT. PT can then use this information to compute second-order information such as the direction of stimuli or whether the stimuli are fast-moving.

\subsection{Motor task model}

In the following, we propose a model for motor tasks (Fig.\ \ref{fig:concmodels}.3). Here we focus on motor timing in which the animals need to time their movement initiation based on sensory stimuli.

According to \cite{Paton2018TheFunctions} \cite{Tsao2022TheDurations}, and \cite{Tallot2020NeuralBrain}, the mechanisms behind motor timing are ramping model or population model-based. Ramping models work with an activity increase over time where actions are produced when activity reaches a threshold \citep{Balc2016ATiming,Kim2017OptogeneticMice,Emmons2020TemporalEnsembles,Bruce2021ExperiencerelatedEncoding}. Population-based models suggest that the timing is encoded by the activity of neural populations \citep{Zhou2022NeuralActivity.,Zhou2022EncodingTradeoffs,Koay2022SequentialInformation}. In general, models can include both ramping activity and encoding of time by neural populations which has been experimentally observed in \cite{Bae2021ParallelNeurons}. In the following, we propose the conceptual model 3 leveraging the local connectivity of IT and PT which is in line with the theory of ramping models and population-based models (Fig.\ \ref{fig:motor_tasks_abstract}).

IT and PT have been implicated in motor timing, with IT being involved in sensory processing and movement preparation and PT being involved in motor timing and movement execution. Work by \cite{Mohan2023CorticalSubnetworks}, \cite{Musall2023PyramidalDecision-making}, and \cite{Bae2021ParallelNeurons} showed through population encoding that these tasks are executed on a population basis. Further, ramping activity by PT has been observed in recent experiments conducted by \cite{Bae2021ParallelNeurons}. Combining all these findings and considering the hierarchical connectivity, we propose that IT sample sensory input (first-order processing) and send this information to PT. PT, in turn, accumulate sensory information over time and save it as an increase in activity. This increase of activity can be interpreted as ramping activity that builds up over time until it reaches a threshold, at which point the motor response is executed (second-order processing). 

In conclusion, we propose a model for motor tasks that is consistent with former ramping and population-based models. IT process sensory information and send it to PT by exploiting the local connectivity, while PT time the motor initiation.

\section{Discussion}
We have proposed conceptual models that are based on the distinct roles of IT and PT. To provide a more complete perspective, we want to discuss studies that challenge the distinct roles of IT and PT. We further propose research directions and future experiments to investigate the functionality of the unidirectional connectivity of IT and PT in information processing. 

\subsection{Overlap of cell-specific temporal roles}
We argued that IT process first-order features whereas PT process second-order features. Despite the theory that IT and PT have distinct temporal roles, the evidence is not entirely conclusive. While studies have shown differences in specialization between the two subtypes, there is also a considerable overlap. \\

In Section \ref{sec:temp}, we highlighted that IT are more selective to orientation (first-order processing) than PT \citep{Kim2015ThreeFunction.}. However, in the same study, they also found that PT showed to be significantly orientation selective. Similarly, \cite{Li2021PhasicDuration} found that both L5b neurons which are most likely PT neurons, and L5a neurons which are most likely IT neurons are selective to tone onset and offset (first-order processing). This suggests that PT do not necessarily become invariant to features processed by IT. 

\cite{Musall2023PyramidalDecision-making} has shown that IT encode first-order features such as stimuli while PT encode second-order features such as choice and movement across cortical brain areas. However, when silencing PT during early audio, late audio, and delay phase in the parietal cortex, there was a larger impairment in performance compared to IT. This suggests that PT play an important role in sensory processing in the parietal cortex. Therefore, PT are relevant for sensory processing in specific cortical areas such as the parietal cortex.

Recent studies by \cite{Mohan2023CorticalSubnetworks} observed that both IT and PT are causally linked to feeding behavior. They found that silencing PT impaired major oral and forelimb movements, including lick and hand lift, while silencing IT during phases of feeding resulted in subtle impairments in finer-scale coordination, such as finger movements during food handling. Thus, both PT and IT were involved in movement execution. Interestingly, silencing IT did not have dramatic effects which does not align with our proposed model where IT input is needed for PT to function. Work by \cite{Park2022MotorClasses} showed further that striatum-projecting IT are relevant for amplitude and speed of movement and pons-projecting PT are relevant for the direction of forelimb movements. Even though IT and PT are relevant for different features, both types are relevant for movement execution indicating another overlap of specialization.

It is worth noting that more subtypes of IT and PT have been identified that perform distinct functions within PT and IT \citep{Economo2018DistinctMovement,Williamson2019ParallelNeurons}. Work that we described in Section \ref{sec:temp} also distinguished between subtypes of PT that either project to striatum or superior colliculus, suggesting that roles within IT or PT are not fully homogeneous \citep{Kim2015ThreeFunction.,Lur2016Projection-SpecificSubnetworks,Ruediger2020LearningCortex}. \\

We acknowledge that there exists an overlap of roles of IT and PT in which IT contribute to second-order processing such as motor execution and PT contribute to first-order processing such as preparatory information processing. However, we also highlighted various studies that support the distinct roles of these cell populations (Section \ref{sec:temp}, Table \ref{tab:functionality}). Considering that cortical L5 plays a crucial role in information processing, in particular temporal processing \citep{Vecchia2020TemporalCortex, Onodera2022TranslaminarLayers}, we believe that the diversity of excitatory cell types in this layer including their local connectivity can contribute to this function. To further investigate this, we want to suggest future research approaches in the following.

\subsection{Future experiments on temporal relevance of IT-PT connection} \label{sec:TempRelITPT}
We have proposed a conceptual model of temporal processing leveraging the local connectivity of IT and PT. Available transgenic mouse lines, genetic tools to label IT and PT, and new optical methods like holographic imaging and photostimulation \citep{Russell2022All-opticalMice} now allow to manipulate IT and PT separately and record from them simultaneously. Although previous experiments have demonstrated the temporal functions of IT and PT, as well as their cellular properties, further experiments are needed to test the functional relevance of the unidirectional connectivity between IT and PT.

To test the conceptual model in sensory discrimination, we suggest different experiments. One possible experiment is to measure the influence of IT on the temporal role of PT in tasks such as direction detection and fast stimuli detection. For example, the temporal specialization (e.g. direction tuning) of PT in animals with silenced IT could be compared to the temporal specialization of PT in a control group with intact IT. If our hypothesis is correct, silenced IT would lead to a decrease in the temporal selectivity of PT compared to the control group.

However, silencing IT would not only target the IT-PT connection but could also target indirect connections from IT to PT. Indirect connections could be those between interneurons and IT/PT cells \citep{Ledderose2021.11.26.469979}. For future research, it would be beneficial to develop experimental methods such as \cite{Vettkotter2022RapidVesicles} that silence synaptic connections to specifically target the IT-PT connection.

Another possible experiment involves the simultaneous recording of IT and PT during a temporal task, with an analysis of correlations of activity regarding their specialization. This can be done by combining two different recombinases for IT and PT with multicolor Calcium indicators \citep{Inoue2019RationalDynamics}. If our hypothesis is correct, the temporal selectivity of PT highly depends on fluctuations in IT activity.

For our model in motor tasks, we presented work by \cite{Musall2023PyramidalDecision-making}, \cite{Mohan2023CorticalSubnetworks} and \cite{Ruediger2020LearningCortex,Nakayama2018Cell-Type-SpecificBehaviors} that silence IT and measure the influence on the behavioral performance. To specifically target the IT-PT connection it would be again beneficial to use methods such as \cite{Vettkotter2022RapidVesicles} that silence synaptic connections.

To sum up, future work could focus on methods that silence synaptic connections to specifically target IT-PT connections and measure the temporal relevance. Other methods would include the simultaneous measurement of IT and PT during a temporal task followed by an analysis of correlations.

\section{Conclusion}
Our work contributes to the question of how the brain processes dynamic input. We presented literature that provides evidence for distinct temporal functionalities of two excitatory neuron subtypes in cortical L5, IT and PT. Our work highlights the significance of these subtypes and their roles in sensory discrimination and motor tasks. We propose that local connectivity might play a crucial role in processing temporal input. More research is needed to explore this connection in terms of functionality which would foster the ongoing investigation of information processing. 

\section{Contributions}
ADV designed research and conceptual models, performed literature review, wrote the paper; EA enriched the content with experimental insights, edited the manuscript, provided feedback; PVA enriched the content with computational insights and edited the manuscript, provided feedback; BFG Project lead, designed research and guided the discussion, provided feedback; KAW Project lead, designed research and conceptual models, provided feedback, edited the manuscript

\section{Acknowledgments}
We thank Dominic Dall’Osto and Pham Huy Nguyen for their helpful comments on the paper. We further thank Benjamin Ehret, Elizabeth Amadei, Laura Sainz Villalba, Martino Sorbaro, Reinhard Loidl, Roman Boehringer, and other members of the Grewe lab for insightful discussions.

\section{Funding}
This work was supported by ETH AI Center to ADV, the European Union’s Horizon 2020
research and innovation programme under the Marie Sklodowska-Curie (101031746) to EA, Swiss National Science Foundation (182539) to PVA, Swiss National Science Foundation (CRSII5-173721, 315230, 189251), ETH Zurich project(ETH-20 19-01), Human Frontiers Science Program (RGY0072/2019) to BFG. 

\section{Author contributions statement}
ADV designed research and conceptual models, performed literature review, wrote the paper; EA enriched the content with experimental insights, edited the manuscript, provided feedback; PVA enriched the content with computational insights and edited the manuscript, provided feedback; BFG Project lead, designed research and guided the discussion, provided feedback; KAW Project lead, designed research and conceptual models, provided feedback, edited the manuscript

\section{Supplementary Material}

\begin{table*}[h!]
\centering
\caption{\textbf{Connectivity between IT and PT.}}
\label{tab:connectivity}
\begin{tabular}{p{4cm}|p{2cm}|p{2.5cm}|p{4cm}}
\textbf{Study} & \textbf{Connectivity} & \textbf{Connection probability} &  \textbf{Brain Area}  \\ \hline \hline
\cite{Brown2009IntracorticalTargets} & IT $\rightarrow$ PT & 19\%  & Mouse Visual Cortex  \\ 
& PT $\rightarrow$ IT & 5\% &   \\ 
& IT $\rightarrow$ IT & 5\% &  \\ 
& PT $\rightarrow$ PT & 7\% &  \\ \hline
\cite{Morishima2006RecurrentCortex} & IT $\rightarrow$ PT & 11  & Rat Frontal Cortex  \\  
& PT $\rightarrow$ IT & 1\% &     \\
& IT $\rightarrow$ IT & 10\% &  \\ \hline
\cite{Kiritani2012HierarchicalCortex} & IT $\rightarrow$ PT & 20\%  & Mouse Motor Cortex  \\ 
& PT $\rightarrow$ IT & 0\% &     \\
& IT $\rightarrow$ IT & 11\% &     \\
& PT $\rightarrow$ PT & 3\% &   \\ \hline
\cite{Campagnola2022LocalNeocortex} & IT $\rightarrow$ PT & 9\%  & Mouse Visual Cortex  \\ 
& PT $\rightarrow$ IT & 0\% & \\ 
& IT $\rightarrow$ IT & 6\% & \\ 
& PT $\rightarrow$ PT & 16\% &  \\ \hline
\end{tabular}

\end{table*}

\newpage

\begin{longtable}{p{0.2cm}|c|c|p{7.1cm}|p{2cm}|p{1.5cm}|p{2.5cm}}
\caption{\textbf{Roles of IT and PT across cortical brain areas in information processing.} This table lists the specializations of IT and PT in sensory discrimination and motor tasks observed in experiments. Abbreviations of brain areas: visual cortex (VC), anterior lateral motor cortex (ALM), auditory cortex (AC), medial prefrontal cortex (mPFC), frontal cortex (FC), motor cortex (MC), parietal cortex (PC), whisker barrel cortex (WC), somatosensory cortex (SC).}
\label{tab:functionality} \\

\hline
\multicolumn{2}{c|}{\shortstack{Temporal \\ Task}} & \shortstack{Cell \\ type} & Related Prediction & \shortstack{Brain \\ area} & \shortstack{Recorded \\ response} & Literature \\ \hline \hline
\endfirsthead

\caption[]{\textbf{Roles of IT and PT across cortical brain areas in information processing (continued).}} \\
\hline
\multicolumn{2}{c|}{\shortstack{Temporal \\ Task}} & \shortstack{Cell \\ type} & Related Prediction & \shortstack{Brain \\ area} & \shortstack{Recorded \\ response} & Literature \\ \hline \hline
\endhead

\hline \multicolumn{7}{r}{\textit{Continued on next page}} \\ \hline
\endfoot

\hline
\endlastfoot

\multirow{10}{*}[-20ex]{\rotatebox[origin=c]{90}{Sensory discrimination tasks}} & \multirow{4}{*}[-3ex]{\rotatebox[origin=c]{90}{\shortstack{Detecting first-\\ order features}}} & \multirow{8}{*}{IT} 
& The activity of IT is tuned to the orientation independent of the direction of moving visual gratings. & Mouse VC & Calcium imaging & \cite{Kim2015ThreeFunction.} \\ \cline{4-7}
  
& & & The activity of IT is sharply tuned to the orientation independent of the direction of moving visual gratings. 
& Mouse VC & Calcium imaging & \cite{Lur2016Projection-SpecificSubnetworks} \\ \cline{4-7}
  
& & & The activity of IT is sharply tuned to the frequency of tones.
& Rat AC & Voltage clamp & \cite{Sun2013Synaptic5} \\ \cline{4-7}

& & & The activity of IT is sharply tuned to the onset and offset of tones.
& Mouse AC & Voltage clamp & \cite{Li2021PhasicDuration} \\ \cline{2-7}

& \multirow{4}{*}[-6ex]{\rotatebox[origin=c]{90}{Direction detection}} & \multirow{11}{*}{PT} 
&  The activity of PT is tuned to the direction of visual gratings.
& Mouse VC & Calcium imaging & \cite{Kim2015ThreeFunction.} \\ \cline{4-7}
  
& & & The activity of PT is broadly tuned to orientation and spatiotemporance of moving visual grating. The activity of PT is sensitive to the direction of low-contrast visual gratings.
& Mouse VC & Calcium imaging & \cite{Lur2016Projection-SpecificSubnetworks} \\ \cline{4-7}
  
& & & The activity of PT is broadly tuned to orientation and spatiotemporance of moving visual gratings.
& Mouse VC & Calcium imaging & \cite{Tang2020LayerBehavior} \\ \cline{4-7}
 
& & & When PT were silenced, the animal's ability to discriminate the direction of low-contrast visual gratings decreased.
& Mouse VC & Behavior & \cite{Ruediger2020LearningCortex} \\ \cline{2-7}

& \multirow{2}{*}[-1.1ex]{\rotatebox[origin=c]{90}{\shortstack{Fast \\ stimuli \\ detection}}} & \multirow{4}{*}{PT} 
& The activity of PT is tuned to the direction of fast-moving visual gratings.
& Mouse VC & Calcium imaging & \cite{Kim2015ThreeFunction.} \\ \cline{4-7}
  
& & & The activity of PT is tuned to short gaps in tones.
& Mouse VC & Tetrodes recording & \cite{Weible2020ADetection} \\ \hline

\multirow{12}{*}[-30ex]{\rotatebox[origin=c]{90}{Motor tasks}} & \multirow{4}{*}[-0ex]{\rotatebox[origin=c]{90}{\shortstack{Sensory process.\\in behav. tasks}}} & \multirow{11}{*}[12ex]{IT} 
& Activity of IT before movement execution encodes stimuli. 
& Mouse mPFC & Calcium imaging & \cite{Bae2021ParallelNeurons} \\ \cline{4-7}
  
& & & Activity of IT encodes stimuli. 
& Mouse PC, FC, AC & Calcium imaging & \multirow{2}{*}{\cite{Musall2023PyramidalDecision-making}} \\ \cline{4-6}
  
& & & When IT were silenced during the sensory stimuli phase, the animals' ability to make a correct choice decreased.
& Mouse FC & Behavior & \\ \cline{4-7}
 
& & & Sensory stimuli activate IT stronger than PT.
& Mouse SC, VC, WC & Calcium imaging & \cite{Mohan2023CorticalSubnetworks} \\ \cline{2-7}

& \multirow{1}{*}[-1ex]{\rotatebox[origin=l]{90}{\shortstack{Motor \\ preparation}}} & \multirow{3}{*}{IT} 
& Activity of IT before movement execution encodes movement type.
& Mouse MC & Calcium imaging & \cite{Currie2022Movement-specificClasses} \\ \cline{4-7}
& & & IT exhibit bilateral selectivity, meaning they can be involved in movements toward both ipsilateral and contralateral sides.
& Mouse ALM & Behavior & \multirow{3}{*}{\cite{Li2015AMovement}} 
\\ \cline{2-7}

& \multirow{2}{*}[-6ex]{\rotatebox[origin=c]{90}{\shortstack{Motor \\ timing}}} & \multirow{6}{*}{PT} 
& Activity of PT encodes time interval before movement execution. In addition, there is ramping activity of PT before movement execution.
& Mouse mPFC & Calcium imaging & \cite{Bae2021ParallelNeurons} \\ \cline{4-7}
  
& & & When PT were silenced at the end of the sensory stimuli phase, the animals' ability to make a correct choice decreased.
& Mouse FC & Behavior & \multirow{3}{*}{\cite{Musall2023PyramidalDecision-making}} \\ \cline{2-6}

& \multirow{3}{*}[-4ex]{\rotatebox[origin=c]{90}{Choice}} & \multirow{7}{*}{PT} 
& Activity of PT encodes choice.
& Mouse PC, FC, AC & Calcium imaging & \\ \cline{4-6}
  
& & & When PT were silenced during movement execution, the animals' ability to make a correct choice decreased.
& Mouse FC & Behavior &  \\ \cline{4-7}
 
& & & When PT were silenced, the animals' ability to adjust their behavior (change their choice) decreased.
& Mouse mPFC & Behavior & \cite{Nakayama2018Cell-Type-SpecificBehaviors} \\ \cline{2-7}

& \multirow{2}{*}[-7ex]{\rotatebox[origin=c]{90}{\shortstack{Motor \\ execution}}} & \multirow{5}{*}{PT} 
& Activity of PT during movement execution encodes movement type.
& Mouse MC & Calcium imaging & \cite{Currie2022Movement-specificClasses} \\ \cline{4-7} 
  
& & &  When PT were silenced during movement execution, the animals' ability to lick and move their forelimbs decreased.
& Mouse SC, VC, WC & Behavior & \cite{Mohan2023CorticalSubnetworks} \\ \cline{4-7} 

& & &  PT exhibit a strong contralateral bias during movement planning, specifically when preparing for contralateral movements.
& Mouse ALM & Behavior & \cite{Li2015AMovement} \\ \hline

\end{longtable}

\newpage

\begin{table*}
\caption{\textbf{Summary of proposed models with unidirectional connection.}}
\label{tab:conceptual_models}
\centering
\begin{tabular}{p{1.8cm}|p{3cm}|p{11cm}} 
 & Name & Description \\ \hline \hline
\textcolor[HTML]{286A8B}{Conceptual Model 1} & Direction Detection & IT detect orientation and send information to PT. PT detect direction using this information. \\ \hline
\textcolor[HTML]{286A8B}{Conceptual Model 2} & Fast stimuli detection & IT detect the frequency and onset and offset of auditory stimuli. IT send this information to PT. PT are more responsive to short gaps in tones using this information. \\ \hline
\textcolor[HTML]{286A8B}{Conceptual Model 3} & Motor task & IT process sensory input and prepare motor execution. IT send this information to PT. PT sum this information up until PT have enough information to time the choice or movement execution. \\ \hline
\end{tabular}
\end{table*}

\begin{table*}[ht]
\caption{\textbf{Sources of visual abstracts figures.} Information about data in figure \ref{fig:sensory_discrimination_abstract} and figure \ref{fig:motor_tasks_abstract}.}
\label{tab:figure_data}
\centering
\begin{tabular}{p{3cm}|p{7cm}|p{5cm}} 
Figure & Description & Source \\ \hline \hline
4.a) (left) Orientation tuning & IT are tuned to orientation. & illustrative abstract based on \cite{Lur2016Projection-SpecificSubnetworks, Kim2015ThreeFunction.} \\ \hline
4.a) (right) Orientation tuning & Figure shows mean orientation selectivity of IT and PT. IT are more orientation selective than PT. & replotted data from \cite{Kim2015ThreeFunction.} figure 6D \\ \hline
4.b) (left) Onset and offset tuning & IT are tuned to onset and offset of tones. & illustrative abstract based on \cite{Li2021PhasicDuration} \\ \hline
4.b) (right) Onset and offset tuning & Figure shows off-responsiveness index (ORI) of L5a and L5b pyramidal
neurons. ORI describes the relative strength of onset and offset responses in an individual cell, with ORI = 1 indicating Off-responsiveness only and ORI = 1 indicating On-responsiveness only. IT and PT are tuned to tone onset and offset. & illustration of experimental results of \cite{Li2021PhasicDuration} figure S2A \\ \hline
4.c) (left) Direction tuning & PT are tuned to direction. & illustrative abstract based on \cite{Lur2016Projection-SpecificSubnetworks,Kim2015ThreeFunction.} \\ \hline
4.c) (right) Direction tuning & Figure shows mean direction selectivity of IT and PT. PT are more direction selective than IT. & replotted data from \cite{Kim2015ThreeFunction.} figure 6D \\ \hline
4.d) (left) Fast stimuli tuning (short gap detection) & PT are tuned to short gaps in tone. & illustrative abstract based on \cite{Weible2020ADetection} \\ \hline
4.d) (right) & Figure describes the median temporal frequency of IT and PT. PT prefer
faster stimuli than IT. & replotted data from \cite{Kim2015ThreeFunction.} figure 6E \\ \hline
5.a) Sensory processing & The figure shows correct decoding accuracy of sensory stimuli by IT and PT. IT
decode visual stimuli better compared to PT. & adapted with permission from \cite{Bae2021ParallelNeurons} figure 3F \\ \hline
5.b) Movement preparation & The figure shows the ability to decode movement type by IT through time. IT decode movement type better before movement initiation than after. & adapted visual abstract with permission from \cite{Currie2022Movement-specificClasses} \\ \hline
5.c) Choice & The figure shows impairment of choice by silencing IT and PT during the experiment. Silencing PT impaires choice stronger during choice compared to silencing IT. & illustration of experimental results from \cite{Musall2023PyramidalDecision-making} figure 8G \\ \hline
5.d) Motor timing & The figure shows decoding accuracy of delay time by IT and PT. PT decode delay time better compared to IT. & adapted with permission from \cite{Bae2021ParallelNeurons} figure 6B \\ \hline
5.e) Movement execution & The figure shows the ability to decode movement type by PT through time. PT decode movement type better after movement initiation than before. & adapted visual abstract with permission from \cite{Currie2022Movement-specificClasses}\\ \hline

\end{tabular}
\end{table*}

\section{References}
\bibliography{references/IT_PT_references,references/mPFC_references,references/temporal_references,references/further_references}

\end{document}